\documentclass{PoS}
\usepackage{amsmath}

\title{Comparisons of Exact Amplitude--Based Resummation Predictions and LHCb Data}

\ShortTitle{Comparisons of Exact Amplitude--Based Resummation Predictions and LHCb Data}

\author{\speaker{Aditi MUKHOPADHYAY}%
\\

        Visiting Scientist, Baylor University\\

        E-mail: \email{aditi.banya@gmail.com}}

\author{B.F.L. Ward\\

        Distinguished Professor of Physics, Baylor University\\

        E-mail: \email{BFL$\_$Ward@baylor.edu}}

\abstract{We present the current status of the comparisons with the respective data of the predictions of our approach of exact amplitude-based resummation in quantum field theory as applied to precision QCD calculations as needed for LHC physics. The agreement between the theoretical predictions and the data exhibited continues to be encouraging.
\\
\begin{center}
BU-HEPP-15-03, Jul., 2015
\end{center}
}

\FullConference{XXIII International Workshop on Deep-Inelastic Scattering\\

                 27 April - May 1 2015\\

                 Dallas, Texas}

\begin{document}

\section{Introduction}
The successful LHC run during 2010 - 2012 has resulted in the accumulation of large samples of data of the Standard Model processes like the heavy gauge boson production and the decay to lepton pairs along with the announcement \cite{1} of the Brout- Englert-Higgs (BEH) \cite{2} boson candidate. The era of prediction of QCD processes at sub-$1\%$ precision tag is upon us. In order to obtain this desired level of accuracy the infrared (IR) improved DGLAP-CS \cite{3, 4} theory \cite{5, 6} realization with HERWIRI1.031 \cite{7} was done by implementing the set of IR improved DGLAP kernels in HERWIG6.5 \cite{8,9}. It has been argued that this process allows better than $1\%$ theoretical precision \cite{10,11}. The comparisons previously done with ATLAS \cite{12} and CMS \cite{13} data are encouraging. In the following we extend this to the LHCb \cite{14} results.

Any precision theory should cover the entire observable phase space for hard processes so that it can completely exploit the entire LHC data. For the single $Z/\gamma^*$ production and the subsequent decay to lepton pairs, the LHCb probes the regime of pseudo rapidity given by $2.0 < \eta < 4.5$ which is different from the regimes probed in the data in Refs. \cite{5,6}. Thus the LHCb provides an oppportunity for a different check. It is to be noted that the cuts on the lepton transverse momenta were similar in all the data.

In what follows we will discuss the comparison of the IR-improved and unimproved parton shower(PS) MC predictions, with the MC@NLO \cite{15} PS/matrix element (ME) matched exact $\mathcal{O}(\alpha_s)$ correction, with the LHCb data on the $Z/\gamma^*$ rapidity, $p_T$ and $\phi^*_\eta$ distributions. The variable $\phi^*_\eta$ was introduced in Refs. \cite{16} in order to overcome the difficulty of measuring $p_T$ spectra for the lower regime of $p_T$. The definition of the variable will be provided in the sections to follow. 

In the next section we will give a brief review of the theory of Exact Amplitude-Based Resummation. In Section. 3 we show the comparison with the LHCb data. We will then discuss the attendant theoretical implications.
\section{Precision QCD for LHC}
We start the discussion in this section with the fully differential representation of a hard LHC scattering process in order to link the experimental results with the theoretical predictions:
\begin{equation}
\small
d\sigma = \sum_{i,j} \int dx_1 dx_2 F_i(x_1) F_j (x_2)d\hat{\sigma}_{res} (x_1 x_2 s)
\label{eq1}
\end{equation}
where $\left\lbrace F_{j} \right\rbrace $ and $d\hat{\sigma}_{res}$ are the parton densities and the reduced hard differential cross section respectively, the later of which has been resummed over all large EW and QCD higher order corrections.

For both the resummation of the reduced cross section and the evolution of the parton densities, the defining formula is identified as
\begin{align}
\small
&d\bar{\sigma}_{res}=e^{SUM_{IR}(QCED)} \sum_{n,m=0}^{\infty} \frac{1}{n!m!} \int \prod_{j_{1}=1}^{n}\frac{d^{3}k_{j_{1}}}{k_{j_{1}}} & \notag \\
&\prod_{j_{2}}^{m} \frac{d^{3}k^{\prime}_{j_{2}}}{k^{\prime}_{j_{2}}} \int \dfrac{d^{4}y}{(2\pi)^{4}} e^{iy.(p_{1}+q_{1}-p_{2}-q_{2}-\sum_{j}k_{j})+D_{QCED}}& \notag \\
&* \tilde{\bar{\beta}}_{n,m}(k_{1},\ldots, k_{n};k_{1}^{\prime},\ldots,k_{m}^{\prime})\dfrac{d^{3}p_{2}}{p_{2}^{0}}\dfrac{d^{3}q_{2}}{q_{2}^{0}}& 
\label{eq2}
\end{align}
where $d\bar{\sigma}_{res}$ is either the reduced differential cross section $d\hat{\sigma}_{res}$ or the evolution rate associated to a DGLAP-CS \cite{3,4} kernel involved in the evolution of the $\left\lbrace F_j\right\rbrace $ and where the new YFS-style \cite{17,18} non Abelian residulas $\tilde{\bar{\beta}}_{n,m}(k_{1},\ldots, k_{n};k_{1}^{\prime},\ldots,k_{m}^{\prime})$ have $n$ hard gluons and $m$ hard photons. Here the final state has been shown with two hard partons with momenta $p_2$, $q_2$ specified for a generic $2f$ final state. The infrared functions $SUM_{IR}(QCED)$ and $D_{QCED}$ are defined as
\begin{eqnarray}
\small
SUM_{IR}(QCED) &=& 2\alpha_s\Re B^{nls}_{QCED}+2\alpha_s\tilde{B}^{nls}_{QCED}\cr
D_{QCED} &=&\int \frac{d^3k}{k^0}\left(e^{-iky}-\theta(K_{max}-k^0)\right){\tilde S}^{nls}_{QCED}
\label{eq3}
\end{eqnarray}
where $K_{max}$ is a dummy parameter and nothing depends on it. The following have been introduced in (\ref{eq3})
\begin{eqnarray}
\small
B^{nls}_{QCED} &\equiv& B^{nls}_{QCD}+{\footnotesize\frac{\alpha}{\alpha_s}}B^{nls}_{QED},\cr
{\tilde B}^{nls}_{QCED}&\equiv& {\tilde B}^{nls}_{QCD}+{\footnotesize\frac{\alpha}{\alpha_s}}{\tilde B}^{nls}_{QED}, \cr
{\tilde S}^{nls}_{QCED}&\equiv& {\tilde S}^{nls}_{QCD}+{\tilde S}^{nls}_{QED}.
\label{eq4}
\end{eqnarray} 
The superscripts $nls$ here denote that the infrared functions are DGLAP-CS synthesized. We stress here that in the formulation of equation (\ref{eq2}) the entire soft gluon phase space is included. The new non-Abelian residuals $\tilde{\bar{\beta}}$ allow rigorous PS/ME matching via their shower substracted analogs. In equation (\ref{eq2}) we make the replacement
\begin{align*}
\small
\tilde{\bar{\beta}}_{n,m} \rightarrow \hat{\tilde{\bar{\beta}}}_{n,m}
\end{align*}
where all effects in the shower associated to the $\left\lbrace F_j\right\rbrace $ are removed from $ \hat{\tilde{\bar{\beta}}}_{n,m}$. The MC@NLO differential cross section can be written as 
\begin{align}
\small
d\sigma_{MC@NLO} &= [B+V+\int(R_{MC}-C)d\Phi_{R}]d\Phi_{B}[\Delta_{MC}(0)+\int (R_{MC}/B)\Delta_{MC}(k_{T}d\Phi_{R})]& \notag \\
&+(R-R_{MC})\Delta_{C}(k_{T})d\Phi_{B}d\Phi_{R}&
\label{eq5}
\end{align}
where the Sudakov form factor, which as usual represents the no emission probablity, is
 \begin{equation}
 \small
\Delta_{MC}(p_{T})=e^{[-\int d\Phi_{R} \frac{R_{MC}(\Phi_{B},\Phi_{R})}{B} \theta (k_{T}(\Phi_{B},\Phi_{R})-p_{T})]},
\label{eq6}
\end{equation}
$B$ is the Born distribution, $V$ is the regularized virtual correction, $C$ is the corresponding counter term required in at exact NLO, $R$ is the respective exact real emission distribution for exact NLO and $R_{MC}$ is the parton shower real emission distribution. We show in Ref. \cite{8} how it is realized via eqn. (\ref{eq2}).

The resulting new resummed kernels, $P^{exp}_{AB}$, yield a new resummed scheme for the PDF's and the reduced cross section:
\begin{align}
\small
\centering
F_{j}, \hat{\sigma} &\rightarrow F_{j}^{\prime}, \sigma^{\prime} for &\\ \notag
P_{gq}(z) \rightarrow P^{exp}_{gq}(z) &= C_{F}F_{YFS}(\gamma_{q})e^{\dfrac{1}{2} \delta_{q}} \dfrac{1+(1-z)^2}{z} z^{\gamma_{q}},  etc.
\label{eq7}
\end{align}
The new scheme gives the same value for $\sigma$ with improved Monte Carlo simulation. Here the YFS infrared factor is $F_{YFS}= e^{-C_{E} a}/ \Gamma(1+a)$ where  $C_{E}$ is the Euler's constant.

The new scheme has improved MC stability. In Herwiri1.031 there is no need for an IR cut-off parameter `$k_0$'. The degrees of freedom below IR cut-offs that are dropped in the usual showers are included in the Herwiri1.031 showers. We note that the difference in the showers starts in $\mathcal{O}(\alpha^2_s)$, the new kernels agree with the old kernels at $\mathcal{O}(\alpha_s)$.
\section{Consistency Checks}
MC HERWIRI1.031\cite{7} is the first realization of the new IR-improved kernels in Herwig6.5 \cite{19} environment. In Refs. \cite{7,8} it was copared with the data of ATLAS and CMS on the single $Z/\gamma^*$ productions and subsequent decay to lepton pairs. 

The MC@NLO/HERWIG6.510  simulations require a PTRMS= 2.2GeV to give good fits to both sets of data whereas the MC@NLO/ HERWIRI1.031 simulations give good fits to both data sets without such an ad hocly large PTRMS [7,8].
 
\begin{figure}
 \centering
\includegraphics[scale=0.33]{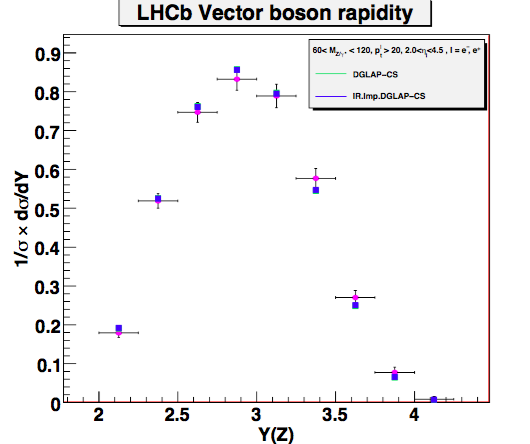}
\caption{Comparison with LHCb data: LHCb rapidity data on ($Z/\gamma^*$)production to $e^+ e^-$ pairs, the circular dots are the data, the green(blue) squares are MC@NLO/HERWIG6.510(PTRMS = 2.2 GeV/c)(MC@NLO/HERWIRI1.031). The green triangles are MC@NLO/HERWIG6.510. These are otherwise untuned theoretical results.}
\label{fig3}
\end{figure}

\begin{figure}
 \centering
\includegraphics[scale=0.33]{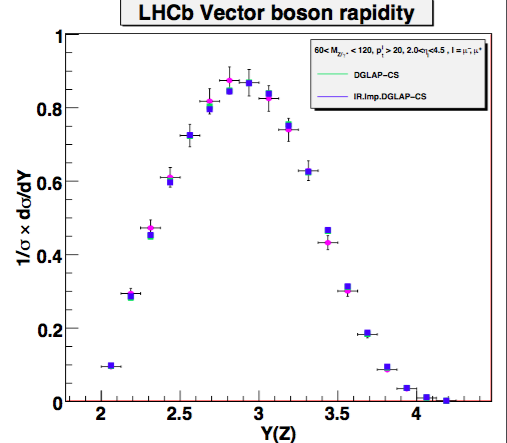}
\caption{Comparison with LHCb data: LHCb rapidity data on ($Z/\gamma^*$) production to bare $\mu^+ \mu^-$ pairs, the circular dots are the data, the green(blue) squares are MC@NLO/HERWIG6.510(PTRMS = 2.2 GeV/c)(MC@NLO/HERWIRI1.031). The green triangles are MC@NLO/HERWIG6.510. These are otherwise untuned theoretical results.}
\label{fig4}
\end{figure}

 We now move on to do the comparison for the LHCb data. Fig.\ref{fig3} shows for the single $Z/\gamma^*$ production at the LHC the comparison between the LHCb rapidity data for the $e^+ e^-$ channel and the MC theory predictions.
Fig.\ref{fig4} shows the comparison with the LHCb rapidity data for the $\mu^+ \mu^-$ channel.
These results should be considered from the perspective of our analysis with the data from ATLAS and CMS in Refs. \cite{7,8}. What we found is that the IR improvement in HERWIRI1.031 allowed it to give a better $\chi^2/d.of$  to the ATLAS and CMS data. It did not need a large intrinsic value of the PTRMS. From already existing models of the proton \cite{20} $PTRMS \simeq 0.4 GeV/c$, which is also what is indicated by the precociousness of Bjorken scaling \cite{21, 22}. The unimproved results from HERWIG6.5 required $PTRMS \cong 2.2 GeV/c$ to get similar $\chi^2/d.o.f$ values for the $p_T$ spectra. However, for the rapidity data such a large value of PTRMS was not required. The LHCb data comparison in Figs. \ref{fig3} and \ref{fig4} show a similar result. The values of $\chi^2/d.o.f$ are 0.746, 0.814, 0.836 for the respective predictions from MC@NLO/HERWIRI1.031, MC@NLO/HERWIG6.5 (PTRMS = 0) and MC@NLO/HERWIG6.5 (PTRMS = 2.2 GeV/c) for the $e^+e^-$ data and are 0.773, 0.555, 0.537 for the respective predictions for the $\mu^+ \mu^-$ data. The values are acceptable for all three calculations.

When we turn to the transverse momentum degrees of freedom, the situation is different from the Refs. \cite{7,8}. We start with the $\phi^*_\eta$ data of LHCb \cite{14}. The definition of this new variable is 
\begin{equation}
\small
\phi^*_\eta = \tan (\phi_{acop}/2) \sqrt{1-\tanh^2(\Delta\eta/2)}
\label{eq8}
\end{equation}
where,
$\Delta\eta = \eta^- - \eta^+$ where $\eta^-$ and $\eta^+$ are the respective negetively and positively charges lepton pseudo rapidities and $\phi_{acop} = \pi - \Delta\phi$ with $\Delta\phi = \phi_1 -\phi_2$ which is the azimuthal angle between the two leptons. It is to be noted here that this variable is not the same as $p_T$ but it is correlated with it.

\begin{figure}
 \centering
\includegraphics[scale=0.33]{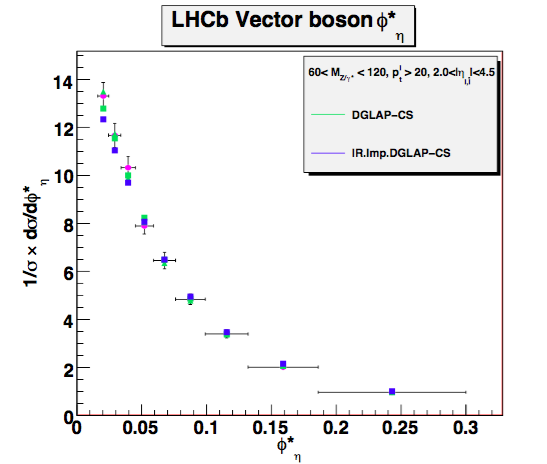}
\caption{Comparison with LHCb data on  $\phi^*_\eta$ for the $\mu^+ \mu^-$ channel in the single $Z/\gamma^*$ production at LHC. The notation is the same as Fig. 2}
\label{fig5}
\end{figure}

Fig. \ref{fig5} shows only the MC@NLO/A results, for A= HERWIG6.5 (PTRMS=0), HERWIG6.5 (PTRMS =2.2GeV/c) and HERWIRI1.031(PTRMS=0) in comparison with the data for the LHCb $\phi^*_\eta$.
We find the respective $\chi^2/d.o.f$ in this case are 1.2, 0.23, 0.35 for the MC@NLO/HERWIRI1.031, MC@NLO/HERWIG6.5 (PTRMS = 0), MC@NLO/HERWIG6.5 (PTRMS = 2.2 GeV/c) simulations. We see that all three calculations give an acceptable fit but the MC@NLO/ HERWIG6.5 (PTRMS=0) gives a mildly better fit than MC@NLO/ HERWIG6.5(PTRMS=2.2 GeV/c). 

\begin{figure}
 \centering
\includegraphics[scale=0.33]{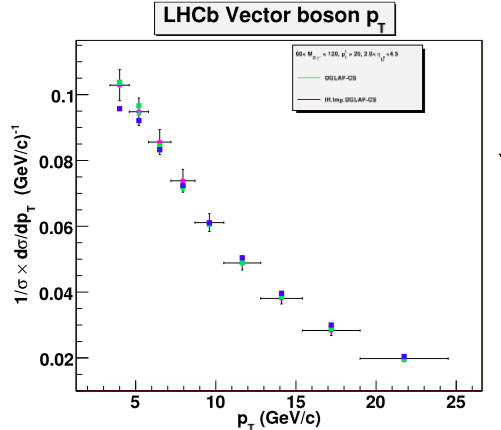}
\caption{Comparison with LHCb data on  $p_T$ for the $\mu^+ \mu^-$ channel in the single $Z/\gamma^*$ production at LHC. The notation is the same as Fig. 2}
\label{fig6}
\end{figure}

In order to be consistent with the comparisons with the CMS and the ATLAS data we turn to compare the $p_T$ spectrum of the LHCb data with the MC theory predictions. Fig. \ref{fig6} shows the corresponding comparison.

The plot shows the comparison between the three MC@NLO/A predictions and the LHCb $p_T$ data, where A = HERWIG6.5(PTRMS = 0), HERWIG6.5(PTRMS = 2.2 GeV/c) and HERWIRI1.031, and as usual we always set PTRMS = 0 in HERWIRI1.031 simulations. The respective $\chi^2/d.o.f$ are 0.183, 0.103, 0.789 respectively. We see that all three calculations give an acceptable fit to the data, with a very mild indication that HERWIG6.5(PTRMS=2.2 GeV/c) gives a better fit than HERWIG6.5(PTRMS=0 GeV/c). We conclude that, when looking at the data on single $Z/\gamma^*$  production at CMS, ATLAS and LHCb , HERWIRI1.031 gives a good fit to the analyzed data without the necessity of ad hocly large intrinsic PTRMS. One of us (A.M.) thanks Dr. Kenneth T. Wilkins, for the kind hospitality of the Baylor College of Arts \& Sciences.

\end{document}